# Hydraulic performance study of hollow adaptive variable pitch tidal energy turbine

Jianmei Chen[1,2+], Chi Xiao[1,2+], Zhuo Wang[1,2], Mingxing Xie[1,2], Wanqiang Zhu[1,2*]

1.*School of Physics, Northeast Normal University, Changchun, 130024, China*
2.*JiLin Provincial Key Laboratory of Advanced Energy Development and Application Innovation, Changchun, 130024, China*

**Abstract**

To address the challenges of bidirectional tidal energy utilization efficiency and operational reliability of tidal turbines under low-flow conditions, this paper presents a novel hollow adaptive variable-pitch tidal energy generator based on symmetric airfoil design.In this document, the hydrodynamic performance of the device is analyzed by CFD method, and the energy capture and thrust load characteristics of the turbine are analyzed and discussed. The simulation results show that the power coefficient of the three-bladed turbine is better than that of the four-bladed and five-bladed turbine in the range of 2-4 tip speed ratios at a pitch angle of 10°, and the optimum power coefficient is 36.8%, which is higher than that of the ordinary axisymmetric wing turbine in terms of its energy acquisition efficiency. Its optimal power coefficient is 36.8%, and its energy efficiency is higher than that of ordinary axial symmetrical wing turbines.

## 1. Introduction

In the era of global fossil energy scarcity, renewable energy is steadily becoming a part of the global energy mix as people are moving towards the goal of global decarbonisation, among which ocean energy is of great exploitation value due to its abundance, availability, low environmental impact and its predictability [1]. Hydraulic turbines have become the key components for capturing tidal energy due to their simple structure, low cost, high efficiency and high reliability.

Usually, hydraulic turbines, in order to adapt to the change flow direction, adopt electric or hydraulic pitch mechanism or integral yaw mechanism, which is easy to realise the active control of the direction, load and the efficiency of energy acquisition [2].

★ Funding: This research was funded by National Key Research and Development Program of China(2024YFE0101200)
*Corresponding author
*Email addresses*: zhuwq773@nenu.edu.cn (Wanqiang Zhu)

In recent years, the self-variable pitch turbine adopting passive current approach has obtained a large amount of real sea condition operation data under the project of bi-directional tidal current support of national ministries and commissions, which has strongly verified the adaptability of this approach to bidirectional tidal current energy. The design of the turbine blades, however, is a key step in the overall design of the turbine, which affects the overall performance of the turbine[3].Since the tidal surge includes two reciprocating processes of high tide and low tide, it is crucial for the tidal energy turbine to have the ability to obtain bidirectional tidal energy, and how to efficiently and reliably utilise the bidirectional tidal energy is still a difficult point in the design of hydraulic turbine. Variable pitch can be divided into two forms: active pitch and passive pitch. The active pitch method mostly adopts electric drive and hydraulic drive to actively control the direction of blade current, energy acquisition and load characteristics, which has become a hotspot in the research of variable pitch. The passive pitch mode is also gaining attention in recent years because it has more advantages in cost, maintenance and reliability compared with the active pitch mode, especially in the complex marine environment.Ma Shun et al [4] proposed a new type of hydraulic pitch actuator, which has a large driving torque, a simple but compact structure, a novel oil circuit, and can meet the needs of the unit's bi-directional operation and power control. Wang Bin et al [5] developed a power-based control scheme and its power-limiting protection strategy for 120kw pitch turbine, which can solve the problem of inaccurate flow measurement and also protect the turbine under limit conditions, and also make full use of bi-directional tidal current energy Currently, horizontal axis hydraulic turbines mostly use the form of blade pitch to obtain bi-directional tidal current energy. In the field of passive pitch turbine, the use of hydrodynamic characteristics of symmetrical airfoil blades to realise passive pitch change (adaptive pitch change) has attracted much attention in recent years because of its high reliability. The current research in this field mainly focuses on the design and hydrodynamic performance of the blades, Chen Jianmei et al [6] studied a new type of blade for adaptive bidirectional flow without originating from pitch change, and analysed the effects of straight and swept blades on pitch change and start-up performance through theory and experiment, Liu Cong et al [7] based on two methods of CFD numerical simulation and physical model experiments for the commonly used adaptive pitch change turbine with The hydrodynamic characteristics of swept back symmetrical airfoil blades were investigated, and the hydrodynamic performances of turbines with different swept back models were compared; Dong Yongjun et al [8] proposed a 300kW adaptive variable pitch HATST, and the hydrodynamic forces and gravitational pitch moments of the blades when the turbine is stationary at rated flow rate were discussed.

Hollow-type tidal energy generators are structurally borrowed from shaftless rim propellers in the marine sector [9]. Hollow type, also known as hollow cross-flow type, is mainly manifested in the centre without obstruction, such as the Open Ventre device of OpenHydeo Company in Ireland. In recent years, there are universities and other units in China gradually designing and researching hollow tidal energy generators, and in 2022, Shanghai Maritime University used the S-wing blade on hollow turbine and carried out a controlled simulation study on axial impellers with the same diameter. The test results show that under the same tip speed ratio, the same size shaftless impeller has higher energy gain coefficient and lower axial thrust; and with the in-crease of shaft diameter ratio, the energy gain coefficient increases and the thrust coefficient decreases. The impeller field flow results show that the shaftless impeller elim-inates the downstream low-speed rotating basin relative to the shafted impeller, further improving the hydrokinetic energy capture efficiency [10]. A similar device is the shaftless rim duct turbine, Songke et al [11] conducted a numerical simulation comparative study between the shaftless rim duct (SRD) turbine and the conventional shafted duct turbine using the computational fluid dynamics (CFD) methodology, and the results of the study showed that, within a certain range of shaft diameter ratios, the shaftless rim duct turbine has a higher output power and a more thrust fluctuation amplitude compared to the shafted duct turbine Compared with the shafted conduit turbine, the fluctuation amplitude is slightly increased, and the fluctuation amplitude further shows an increasing trend with the decrease of the impeller shaft diameter ratio.

To enhance the adaptability of tidal current power generation devices to low-current open-sea environments and to efficiently and reliably harness bidirectional tidal energy, this paper proposes a novel hollow adaptive pitch-variable tidal current power generation device. Unlike conventional axial horizontal axis power generation devices, this design features a hollow, obstruction-free central structure, enabling consistent start-up current speeds in both forward and reverse tidal flows. Additionally, the use of symmetrical airfoil blades simplifies the overall structure, improving reliability, and allows the blade pitching process to be achieved solely through the action of water currents. The structure of the device is more simple and re-liable, and at the same time, it can realise the blade pitching process only under the action of water flow. The device is an internal rotor structure, the motor rotor and impeller are integrated, the generator is installed in the stator inside, when the incoming flow, the trailing edge of the blade will be deflected backward around the axis of rotation, thus changing the blade mounting angle up to the maximum limit value, when the current flow rate reaches the positive mounting angle of the starting speed, the blade drives the rotor to start clockwise rotation, so as to capture the tidal current energy. This paper focuses on the hydrodynamic performance of the impeller section.

## 2. Adaptive Variable Pitch Structure
### 2.1 Principle of adaptive variable pitch

Adaptive pitch change leverages the hydrodynamic characteristics of symmetric airfoil blades, enabling the blades to achieve passive pitch adjustment solely through the action of water flow, without the need for additional energy input. [12]. A schematic diagram of the vane pitch based on a symmetric airfoil blade is shown in Fig. 1, where P is the intersection of the blade pitch axis with the current vane, and C is the hydrodynamic centre point of the vane. Symmetric airfoil pressure centre is usually located in the distance from the leading edge of the rear quarter chord length, the centre of the pitch axis of the leaf element needs to be set between the hydrodynamic centre and the leading edge, when the forward current impacts on the blade will be relative to the centre of the pitch axis to form a clockwise direction of the deflection torque, when the deflection torque is large enough to overcome the frictional resistance torque of the leaf root rotating axis, the blade will be around the pitch axis to complete the passive pitch pitch; when the reverse current impacts on the blade, the blade in the counter-clockwise direction, the pitch axis is in the counter-clockwise direction, the pitch axis is in the counter-clockwise direction, the pitch axis is in the counter-clockwise direction. When the reverse current hits the blade, the blade completes the passive pitch change under the action of the counterclockwise deflection torque [13].

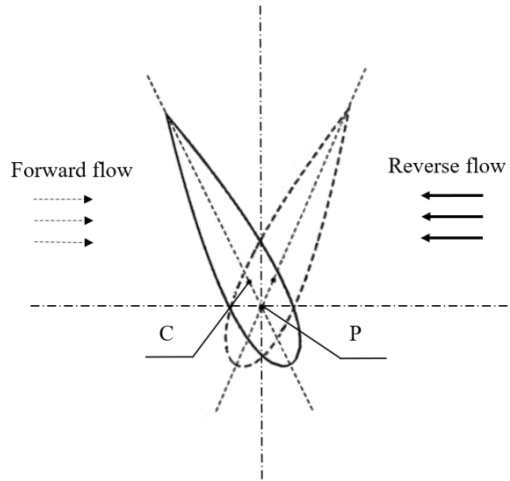

**Figure 1.** Schematic diagram of blade element variable pitch.

2.2 Principle of operation of hollow adaptive pitch turbine

As the forward flow velocity increases from zero, the trailing edge of the blades begins to deflect backward around the rotating axis when the hydrodynamic moment acting on the blades exceeds the combined resistance moment from both the frictional force at the root rotating axis and the fluid resistance on the blades. This deflection gradually adjusts the blade mounting angle until it reaches its maximum limit.[14]. As shown in Fig. 2, when the water flow velocity reaches the starting speed for the positive mounting angle, the blade starts to rotate clockwise, thus capturing the tidal energy.

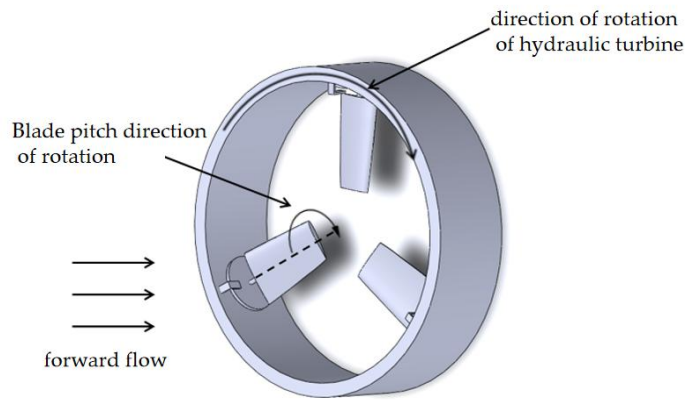

**Figure 2.** Working Principle Diagram for Forward Incoming Flow

As illustrated in Fig. 3, the schematic diagram depicts the operating principle under reverse incoming flow. The behavior of the turbine under reverse flow conditions is identical to its operation under forward flow. This design enables the turbine blades to achieve passive pitch adjustment solely through the action of the current, allowing the system to efficiently adapt to bidirectional flows.

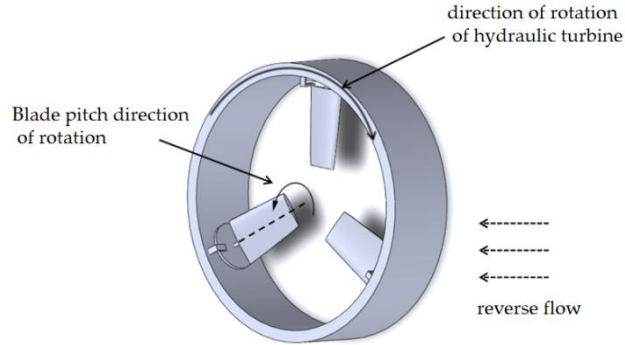

**Figure 3.** Working Principle Diagram for Reverse Incoming Flow

**3. Blade design and modelling**

3.1 Airfoil Selection

In this study, the blade of the initial model is designed using the modified blade element momentum theory [15]. The NACA0012 airfoil is selected based on its aerodynamic characteristics, and the blade design parameters are optimized to maximize the power coefficient of a single blade section. A program is developed using MATLAB software to calculate the chord lengths and other parameters that achieve the maximum power coefficient for each cross-section of the blade. Since this paper is based on the existing conditions in the laboratory to design the proposed output power of the turbine is 0.32W, the rated flow rate is 0.305m/s, the predicted power coefficient is 0.30, to get the corresponding parameters of the turbine, and then get the chord length of the blades, the parameters of the chord length of the various cross-sections are shown in Table 1. The 3D and physical drawings of the tidal energy generator are shown in Fig. 4.

**Table 1.** Parameters of each cross-section of the blade

| Wing span position $r/R$ | Blade chord length $c$/mm | Wing span position $r/R$ | Blade chord length $c$/mm |
|---|---|---|---|
| 0.2$R$ | 42.18 | 0.65$R$ | 51.47 |
| 0.29$R$ | 44.81 | 0.73$R$ | 53.12 |
| 0.38$R$ | 46.67 | 0.82$R$ | 54.26 |
| 0.47$R$ | 48.05 | 0.91$R$ | 55.73 |
| 0.56$R$ | 49.88 | 1$R$ | 57.03 |

The chord lengths used were imported into Profili to get the 3D coordinates of the used airfoil, and then imported into Solidworks to build the 3D model of the blade.

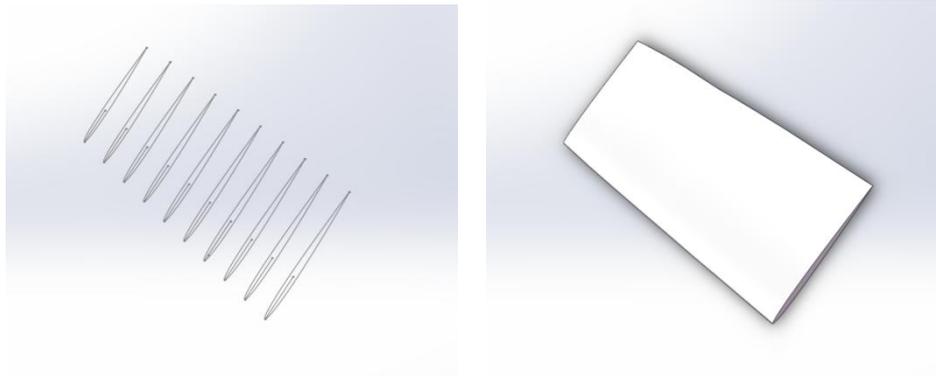

**Figure 4.** 3D view of the blade

3.2 Effect of shaft diameter ratio on energy efficiency

To enhance the energy capture efficiency of this hollow tidal current turbine, simulation studies revealed that adjusting the blade length while maintaining a fixed rotor diameter significantly impacts the efficiency of energy extraction. [10]. Therefore, in this paper, the hydrodynamic performance of three hollow impellers with different shaft diameter ratios will be investigated, where the shaft diameter ratio is defined as the ratio of the distance of the blades from the centre, r, to the distance of the rotor from the centre, R, i.e., r/R. The change in shaft diameter ratio is achieved by varying the actual radius of the outer rotor, R. Table 2 shows the table of values of shaft diameter ratio.

**Table 2.** Table of values for shaft diameter ratio

| Shaft diameter ratio r/R | R (mm) |
|---|---|
| 0.7 | 135 |
| 0.8 | 117.2 |
| 0.9 | 104.18 |

The hydrodynamic performance of a tidal current energy turbine can be characterised by its energy capture efficiency at different tip speed ratios, which is given by the following equation:

$$\lambda = \frac{2\pi R n}{v} = \frac{wR}{v} \qquad (1)$$

Where w is the rotational angular velocity of the impeller at a stream velocity of v, $w = \frac{2\pi}{60} n$ ; n is the rotational speed.

The variation of energy yield efficiency with tip speed ratio for a three-bladed hollow tidal current energy generator at the optimum pitch angle of 10o is shown in Fig. 5 for the selection of shaft diameter ratio.

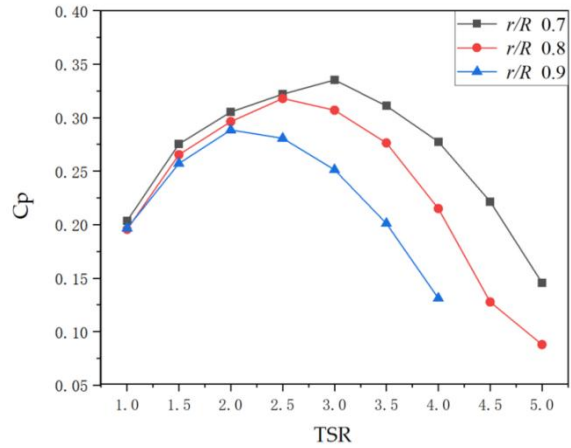

Figure 5. Comparison of power coefficients for different shaft diameter ratios

As can be seen from the above figure, the impeller energy gain coefficient increases with the increase of the tip speed ratio, when the tip speed ratio is in the range of 2-3, the power coefficient of the impeller reaches the maximum value, but with the gradual increase of the tip speed ratio, the impeller's energy gain efficiency begins to decrease. When the tip speed ratio is 4-5, the impeller efficiency is extremely low, especially the shaft diameter ratio of 0.9 and shaft diameter ratio of 0.8, and the lowest energy efficiency is only 7%, which shows that the best tip speed ratio of this device are in the range of 2-3, which is in line with the normal operation of the low-flow hydraulic turbine. Under different tip speed ratios, the impeller energy efficiency is the best when the shaft diameter ratio is 0.7. Therefore, the shaft diameter ratio of 0.7 is chosen for the subsequent simulation of the device and the fabrication of the solid model of the device for the research and test.

## 4. Impeller hydrodynamic performance study

4.1 Meshing

Firstly, the 3D model of the turbine and fluid domain is established, and the design of the computational domain with a length of 10D, a width of 3D, and a height of 4D for the fluid computational domain is shown in Figure 6.

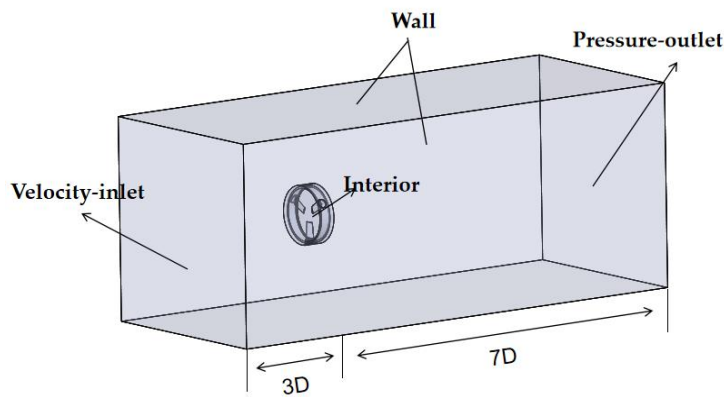

Figure 6. Computational domain

Numerical modelling and simulation of the hydraulic turbine model based on Computational Fluid Dynamics (CFD) methodology is carried out using the pre-processing software ICEM CFD for mesh generation and optimization [16]. Before mesh division, we first establish Part and name each part of the computational domain, which are the inner basin, outer basin, wall, and rotor and blade, and then set the mesh size for each part, in order to better simulate the flow effect, the inner basin in this computational domain uses the wall function method to get the height of the first boundary layer of the model to be 0.23mm through the formula of Y+, and the number of boundary layers is set to be 0.23mm according to computer Resource configuration set the number of boundary layers for 4 layers, the height of the boundary layer compartments follow an exponential distribution, the coefficient is 1.2, set the grid division method for the tetrahedral mesh method, the obtained grid is shown in Fig. 7.

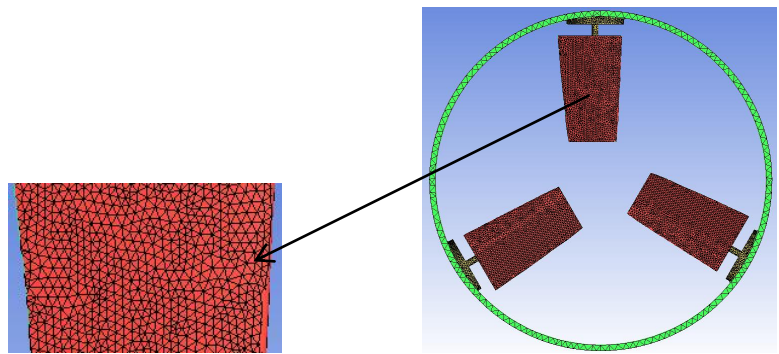

Figure 7. Boundary conditions

4.2 Grid-independent verification

The accuracy and economy of the scheme should be considered when designing a tidal energy turbine. The number of nodes generated by the above scheme is about 600,000 and the total number of grids is about 3.42 million. Grid independence test is required. For the model, five grid encryption methods are discussed and the grids are encrypted in the inner and outer domains, respectively. The results were obtained for different number of nodes and number of grids as shown in Table 3.

Table 3. Grid independence verification

| Programme | Number of Nodes | Total Number of Grids | Power Coefficien |
|---|---|---|---|
| 1 | 285494 | 1615728 | 0.2774 |
| 2 | 453130 | 2568537 | 0.2802 |
| 3 | 607711 | 3428975 | 0.2889 |
| 4 | 655496 | 3704586 | 0.2651 |
| 5 | 712796 | 4040672 | 0.2598 |

From Table 3, it is evident that the power coefficient remains relatively stable even as the number of grids increases exponentially. This suggests that a higher total number of grids does not necessarily lead to an improvement in energy capture efficiency. Within a limited range of grid density, the denser the grid, the higher the computational accuracy. The comparison between Scenario 5 and Scenario 4 shows that beyond this limit, further densification cannot effectively improve the accuracy. When the grid density limit is exceeded, the number of different grids does not affect the simulation results significantly. Overall, the mesh in Scheme 3 achieves more satisfactory prediction results. Considering the computational resource allocation and simulation efficiency, the more grids, the longer the computation time, so the subsequent simulation can choose scheme 3.

4.3 Analysis of simulation results

The direct influence of the force on the hydraulic turbine is the thrust coefficient Ct, which is expressed as:

$$c_t = \frac{T}{\frac{1}{2}\rho V_1^2 \pi R^2}$$
(2)

The formula: T is the axial thrust force on the hydraulic turbine.

The energy capture coefficient Cp is an important parameter for evaluating the hydraulic turbine and its expression as:

$$C_P = \frac{P}{\frac{1}{2}\rho V_1^3 \pi R^2}$$
(3)

The starting torque is the torque of the static blade under the action of water flow. From literature [8] the relationship between pitch angle and blade starting torque can be calculated by equation (4):

$$Q = \frac{\rho v^2 R}{2} \int_0^1 l(x) \sin(2\theta(x)) dx$$
(4)

Equation, $Q$ - torque (starting torque) applied before blade rotation; $\rho$- fluid density; V-fluid velocity; $R$-blade radius; l(x) -blade chord length distribution; $\theta(x)$-blade pitch angle distribution; x-position of leaf vein as a percentage of blade length as a percentage (from leaf root to tip 0 to 1). From equation (4), when the chord length distribution is determined, the starting torque is positively correlated with the blade pitch angle in the range of 45°. For no torsion angle the larger the blade pitch angle, the easier the impeller is to start [17].

The starting torque is the torque of the static blade under the action of the current. Table 4 gives the model starting torque corresponding to the change in velocity range from 0.2 m/s to 0.5 m/s for different pitch angles in forward incoming flow. From the data, it is observed that the starting torque of the model increases with the increase in flow velocity. The starting torque also increases gradually as the pitch angle increases. And the turbine still has a more satisfactory torque when the incoming flow velocity is 0.2 m/s, which verifies that the hollow turbine can still start automatically when the incoming flow is low.

Table 4. Starting moments at different pitch angles for forward incoming flow

| Flow Rate (m/s) | β=-8° | β=-10° | β=-12° | β=-14° |
|---|---|---|---|---|
| 0.2 | 0.005811 | 0.007865 | 0.009835 | 0.011375 |
| 0.3 | 0.011245 | 0.014765 | 0.017925 | 0.027649 |
| 0.4 | 0.020137 | 0.032746 | 0.039735 | 0.041728 |
| 0.5 | 0.036149 | 0.040976 | 0.051287 | 0.061642 |

In order to verify that the device can adapt to bidirectional currents more efficiently, Table 5 gives the model starting moments corresponding to the changes in the velocity range from 0.2 m/s to 0.5 m/s for different pitch angles in the reverse incoming flow. From the table, it can be seen that the difference between the starting moments of the device in reverse incoming flow and in forward incoming flow is less than 1%, which verifies the design purpose of the device and the reasonableness of the device.

Table 5. Starting moments at different pitch angles for reverse incoming flow

| Flow Rate (m/s) | β=8° | β=10° | β=12° | β=14° |
|---|---|---|---|---|
| 0.2 | 0.005236 | 0.007026 | 0.009035 | 0.009257 |
| 0.3 | 0.009658 | 0.012587 | 0.016842 | 0.020658 |
| 0.4 | 0.019785 | 0.029687 | 0.032541 | 0.036987 |
| 0.5 | 0.030257 | 0.038542 | 0.047259 | 0.059325 |

4.4 Power coefficien

Figure 8 presents the variation of power coefficient with tip speed ratio, pitch angle for a hollow symmetrical airfoil blade turbine at a flow rate of 0.3 m/s and number of blades 3.

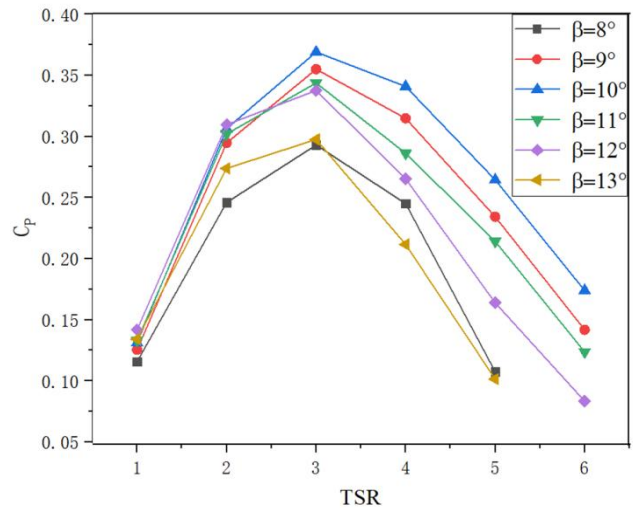

Figure 8. Power coefficient of three-bladed hollow turbine

As can be observed in the figure, when the pitch angle of the blades is certain, for the three-bladed hollow-type variable pitch turbine, the energy gain coefficient is relatively low at low tip speed ratio, while the energy gain coefficient is relatively high at high tip speed ratio; within the range of 8°-14° pitch angle of the blades, the power coefficient of the turbine shows the trend of increasing and then decreasing with the increase of the pitch angle when the pitch angle of the blades is certain; the range of the tip speed ratio for the stable operation also decreases with the increase of the pitch angle; the optimum tip speed ratio at the corresponding pitch angle also decreases with the increase of the pitch angle, and the optimum gain ratio is 10°, and the optimum gain ratio at this time is also 10°. The range of stable working tip speed ratio also decreases with the increase of pitch angle; the optimal tip speed ratio under the corresponding pitch angle also decreases with the increase of pitch angle, and the optimal pitch angle is 10°, at which time the optimal energy-acquiring tip speed ratio is 3, and the optimal power coefficient can reach 36.8%.

Figure 9 displays the variation of power coefficient with tip speed ratio, pitch angle for a symmetrical wing blade turbine at a flow rate of 0.3 m/s and number of blades 4.

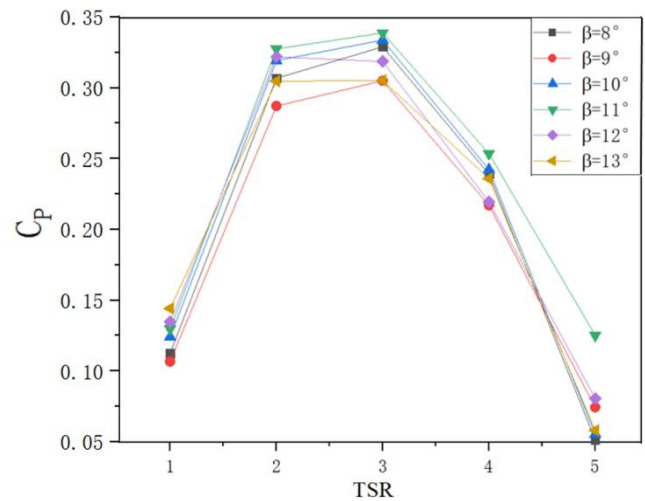

Figure 9. Power coefficient of four blade turbine

The diagram indicates that when the pitch angle of the blades is certain, the power gain coefficient of the four-bladed turbine increases with the increase of the tip speed ratio, and the power coefficient decreases sharply after the optimal tip speed ratio, and it can be found that the power coefficient of the turbine has been reduced to less than 10% when the tip speed ratio is greater than 4.5, so it can be seen that the turbine of the symmetrical airfoil with four-bladed is suitable for working under the condition of lower tip speed ratio, and can obtain a better power coefficient at the low rotational speed. It can be seen that the four-bladed symmetric blade hydraulic turbine is suitable to work under lower tip speed ratio and can obtain better power coefficient at lower rotational speed; in the range of 8°-13° blade pitch angle, the power coefficient of the hydraulic turbine increases with the increase of the specific tip speed ratio, and the power coefficient of the hydraulic turbine does not decrease much in the range of the low tip speed ratio after exceeding the optimal pitch angle, and the power coefficient of the hydraulic turbine decreases in the range of 10% at the high tip speed ratio, and the range of the suitable tip speed ratio decreases gradually with the increase of the pitch angle up to 11°, and the optimal tip speed ratio under the same pitch angle increases to 11°. The optimum tip speed ratio at the same pitch angle also decreases gradually after the pitch angle of 10°,and the maximum energy gain coefficient reaches 33.2% at the pitch angle of 10° and the optimum tip speed ratio of 3.

Figure 10 demonstrates the variation of power coefficient with tip speed ratio and pitch angle for a symmetrical airfoil blade turbine at a flow rate of 0.3 m/s and a number of blades of five.

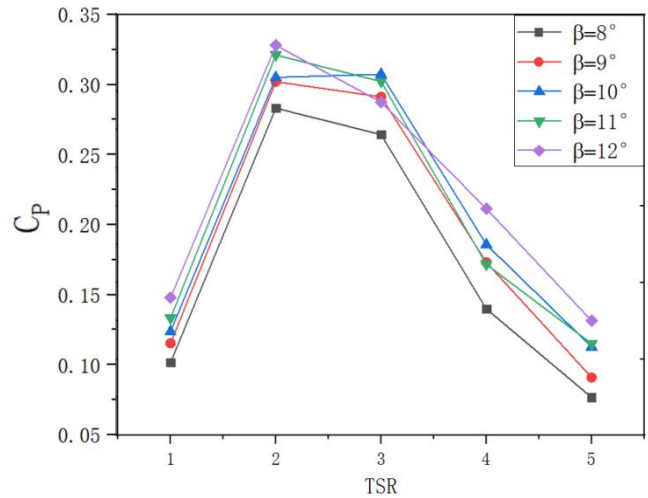

Figure 10 .Power coefficient of five blade turbine

The images reveals that the pitch angle of the blades is certain, the energy gain coefficient of the five-bladed turbine increases with the increase of the tip speed ratio, and the power coefficient gradually decreases after the optimal tip speed ratio is exceeded, unlike the three- and four-bladed devices, the optimal tip speed ratio of the five-bladed device is 2, and due to the lower number of the blades, the device has a relatively good energy gain efficiency under the tip speed ratio of 2, and the device's energy gain efficiency decreases to below 20% at tip speed ratios from 8° to 13°. When the tip speed ratio is greater than 3, the energy efficiency of the device has been reduced to less than 20% at pitch angles from 8° to 13°, which shows that the hollow hydraulic turbine with five symmetrical airfoil blades is suitable for working at lower tip speed ratios, and can obtain better power coefficients at lower rotational speeds;In the low tip speed ratio both when the tip speed ratio is 1, the power coefficient of the device increases with the increase of the pitch angle, and after that, with the increase of the tip speed ratio, its corresponding optimal pitch angle changes, and the maximum energy gain coefficient reaches 32.5% when the pitch angle is 12 degrees and the optimal tip speed ratio is 2.

The turbine with symmetrical airfoil blades with three kinds of blade numbers is shown that the power coefficient of the turbine increases and then decreases with the increase of the tip speed ratio under a certain pitch angle, and the tip speed ratios to reach the optimal power coefficient are different for different blade numbers, and the best energy acquisition characteristics are reached when the tip speed ratio is 3 for the three and four blades, and the highest power coefficient of 30% for the three blades, and the best power coefficient of 26% for the four blades. The maximum power coefficient of three blades is 29% and the optimum power coefficient of four blades is 26%. And because of the four blades in the tip speed ratio of 5, the lowest energy efficiency is only 1%, so the three blades in the high tip speed ratio is more stable compared to the four blades; five blades in the tip speed ratio of 2 to achieve the best energy efficiency of the best energy efficiency of 28%, but because of the five blades of the energy efficiency of the different tip speed ratio is very unstable, only in the tip speed ratio of 2 and 3 when the energy efficiency of the 20%, and the three blades in the pitch angle of 10 °, in the 2-5 at the best power coefficient of 26%, four blades the best power coefficient of 26%. At pitch angle 10°, the energy harvesting efficiency is greater than or equal to 20 per cent at tip ratios ranging from 2 to 5.

4.5 Thrust Coefficient

Because of the complex and variable offshore conditions, the operational reliability of tidal current energy generators is extremely important. Figure 11 gives the variation of thrust coefficients with tip speed ratio and pitch angle for a symmetrical wing blade turbine with a flow velocity of 0.3 m/s and a number of blades of three blades.

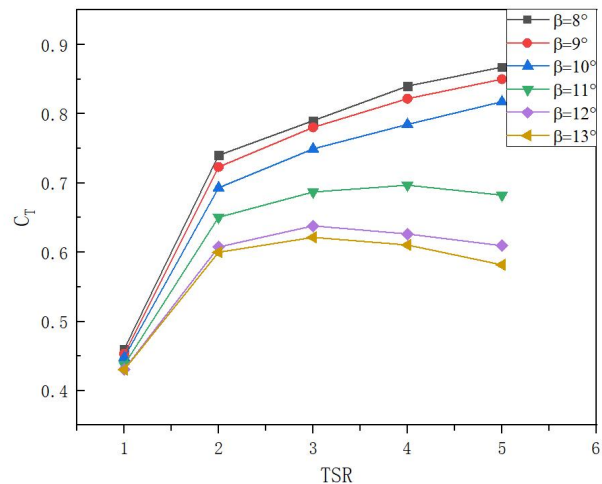

Figure 11. Thrust coefficients for a three-bladed turbine

From the analysis of the above figure, it can be seen that the thrust coefficient CT of the hydraulic turbine decreases with the increase of the pitch angle. Due to the increase of the pitch angle, the actual flow area of the blades decreases, which leads to a reduction of the thrust load acting on the blades. When the pitch angle is less than or equal to 10°, the tip speed ratio 1-5 range, the thrust coefficient increases with the increase of the tip speed ratio When the tip speed ratio exceeds the optimal energy-acquiring tip speed ratio at this pitch angle, the increase of the thrust coefficient is small; when the pitch angle is greater than 10°, the thrust coefficient of the hydraulic turbine increases and then decreases with the increase of the tip speed ratio.

Figure 12 indicates the variation of thrust coefficient with tip speed ratio and pitch angle for a symmetrical airfoil blade turbine with a flow velocity of 0.3 m/s and a number of blades of four blades.

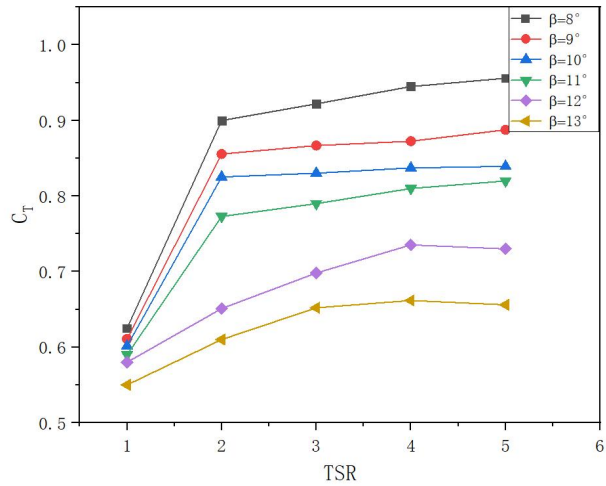

Figure 12. Thrust coefficients for four-bladed turbine

From the figure, it can be observed that the thrust coefficient of the four-bladed turbine gradually increases with higher tip speed ratios and pitch angles within the range of 8° to 13° and tip speed ratios from 1 to 5. However, the rate of increase diminishes as the tip speed ratio rises.

Figure 13 presents the variation of thrust coefficient with tip speed ratio and pitch angle for a symmetrical airfoil blade turbine at a flow rate of 0.3 m/s and a number of blades of five blades.

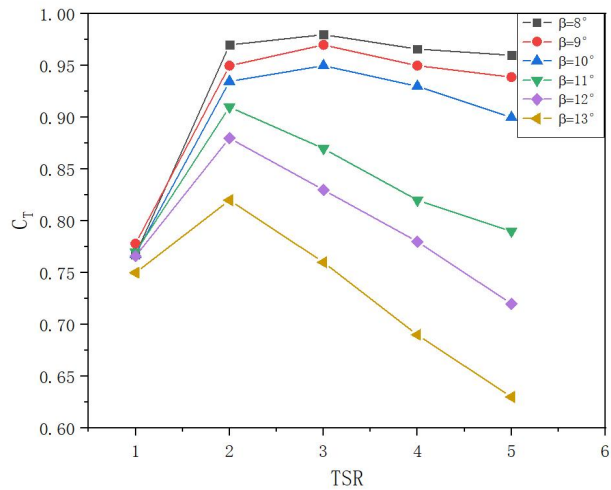

Figure 13. Thrust coefficients for five-bladed turbine

As you can surmise from the graph, the thrust coefficient of the five-bladed hydraulic turbine shows a tendency of increasing first and then decreasing with the increase of the tip speed ratio, and with the increase of the pitch angle as well as the increase of the tip speed ratio, the decrease of its thrust coefficient increases after reaching the maximum thrust coefficient.

Combining the above results, the thrust coefficient of the hydraulic turbine decreases with the increase of the pitch angle when the tip speed ratio is certain; the thrust coefficient increases with the increase of the tip speed ratio at a smaller pitch angle, and the increase of the thrust coefficient is smaller when the tip speed ratio exceeds the optimal energy-acquiring tip speed ratio at this pitch angle; the thrust coefficient increases with the change of the tip speed ratio and decreases with the change of the tip speed ratio at relatively large pitch angles, and the thrust coefficient increases with the increase of the tip speed ratio and pitch angle at the same tip speed ratio and pitch angle. Under the same tip speed ratio and pitch angle, the more the number of blades, the larger the thrust coefficient, and it can be seen that the thrust coefficient of five blades is the largest, and the thrust coefficient of three kinds of hydraulic turbines with three numbers of blades is the smallest of 43.25% for three blades, and the difference between the thrust coefficients of four blades and five blades is smaller, respectively, 55% and 63%, at the pitch angle and the tip speed ratio of the turbine with the optimal power coefficients.

Based on the above study, this paper selects the three-bladed device with considerable and stable energy yield efficiency and the smallest thrust coefficient for experimental and other simulation studies.

4.6 Impeller axial flow velocity analysis

The rotating blades of tidal current energy turbine will have a certain blocking effect on the surrounding water movement, which affects the normal flow path of the water and has a great influence on the arrangement of the turbine [18]. Therefore, it is necessary to study the wake field of tidal current energy turbine. Figure 14 illustrates the wake distribution of a three-bladed hollow turbine at a flow velocity of 0.3 m/s and a tip-speed ratio of 3 with different pitch angles.

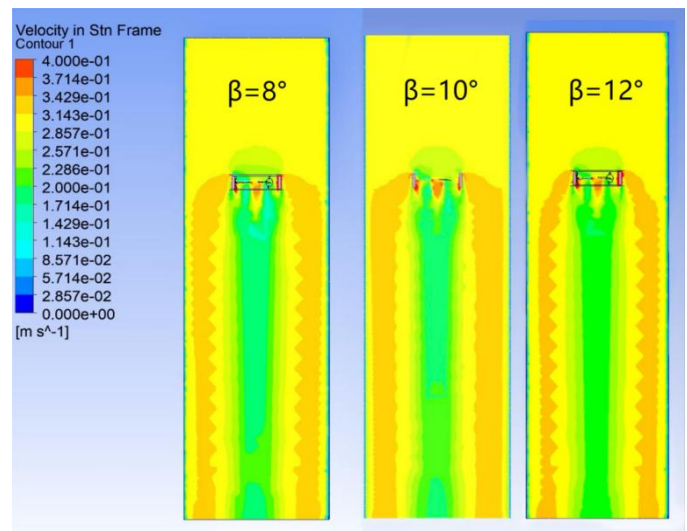

Figure14．Turbine wake distribution at different pitch angles

From the diagram, when the impeller rotates under the impact of fluid, there is a pressure difference between its front and back. Unlike the traditional shafted turbine, the hollow turbine has no support structure and central axis in the centre, so in a sweep cycle, the central sweep notch always exists and has the tendency of alternating existence with the rotation of the blades, so there is no presence of the low velocity band of the shafted impeller, and after the fluid flows through the shaftless impeller, because of the existence of the pressure difference between the front and the rear, the fluid is squeezed to form a flow velocity band of the rear with a uniform relative velocity. Under the same tip speed ratio, the flow velocity at the same axial distance behind the turbine increases with the increase of pitch angle, which is caused by the different kinetic energy obtained from the water under different pitch angles, and at the same time, the larger the pitch angle is, the relative reduction of the turbine's flow area, and the blocking effect of the water flow will also be reduced.

Figure15 gives the variation of wake velocity distribution with tip-speed ratio for a three-bladed hollow turbine at a flow velocity of 0.3 m/s with a pitch angle of 10°.

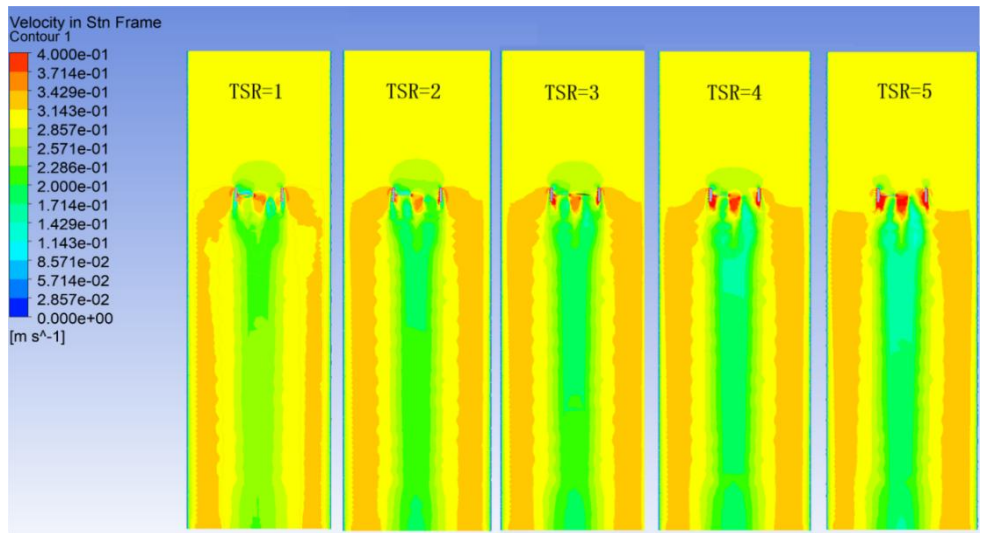

Figure 15． Cloud figure of turbine wake distribution under different tip speed ratios

Since there is no obstruction in the middle of the hollow turbine, the rear of the turbine is a high flow velocity area, and the high flow velocity area increases with the increase of tip speed ratio, and at the same time, with the increase of the tip speed ratio, the energy obtaining efficiency of the turbine increases, which results in the increase of the kinetic energy absorbed by the turbine, so the wake velocity decreases gradually. Because of the optimal tip speed ratio when the turbine is working, the turbine's ability to extract kinetic energy from the water is better at this time, so the recovery of the wake velocity at the rear is relatively slow, and the wake velocity at the same axial distance at the rear of the turbine shows a decreasing trend with the increase of the tip speed ratio. Due to the disturbance of the blade tip, the flow velocity outside the diameter of the turbine is increased, and the influence range also shows a tendency of increasing and then decreasing with the increase of the tip speed ratio.

4.7 Turbine Blade Pressure Distribution

The turbine blades will be subjected to a large on-stream load when the water flow acts on the impeller, and the pressure load is closely related to the static pressure distribution of the blades, which is inextricably linked to the reliability of the device, so it is crucial to analyse the static pressure distribution of the blades under different working conditions.

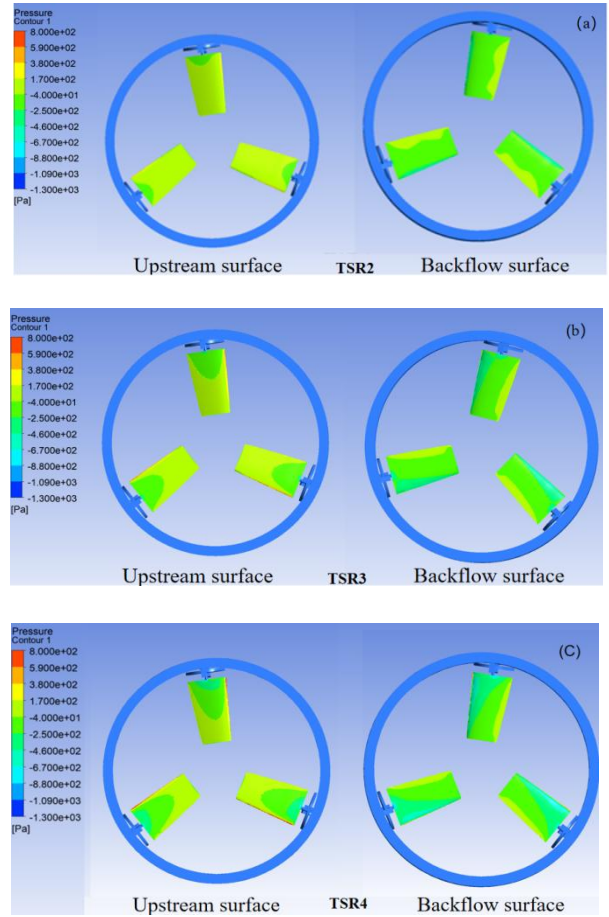

Figure 16．Cloud diagram of blade surface pressure distribution under different tip speed ratios
(a) TSR is 2; (b) TSR is 3; (c) TSR is 4.

Figure 16 illustrates the distribution of static pressure on the onward face of the blades of this hollow turbine at a flow velocity of 0.3 m/s and a pitch angle of 10° for different tip speed ratios.From the blade surface pressure distribution diagram, the high pressure on the blade head-on surface is mainly concentrated in the blade root area, with the increase of the tip speed ratio, the size of the pressure on the blade root and the range of increase also increases, resulting in a larger torque and thrust acting on the blade, thus increasing the energy acquisition and load characteristics of the turbine; when the tip speed ratio is too high, the increase in the magnitude of such a pressure difference decreases, and the negative pressure area of the head-on surface also increases. When the tip speed ratio is too high, this pressure difference decreases, and the negative pressure area on the headwater surface also gradually expands to the trailing edge, which leads to the reduction of the torque and thrust of the blades, which is in line with the results of the hydrodynamic simulation in this paper and the existing research, and the energy gain characteristic increases and then decreases with the increase of the tip speed ratio, and the load characteristic increases and then gradually stabilises or gradually decreases.

According to the results of the cloud diagram and its existing hydrodynamic theory, it can be seen that the static pressure at different cross sections of the same blade is not the same [20], Fig. 17-19 give the same pitch angle in different tip speed ratio when the blade at different cross sections of the static pressure cloud diagram. From the figure analysis, when the pitch angle is certain, the closer the blade cross-section is to the tip, the positive pressure on the in-flow surface gradually increases, and the negative pressure on the back-flow surface gradually decreases, indicating that the power provided for the blade is concentrated around the tip. At the same blade cross-section position, the positive pressure on the inlet surface of the blade increases with the increase of the tip speed ratio, while the negative pressure on the backflow surface decreases gradually, so the hydraulic turbine will lead to an increase in thrust with the increase of the tip speed ratio.

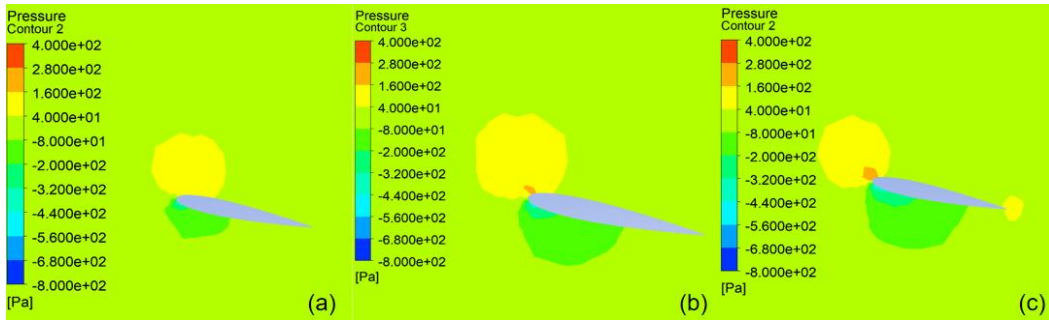

Figure 17. Blade 0.42R section pressure at 12°pitch angle (a) TSR = 2 (b) TSR = 3 (c) TSR = 4

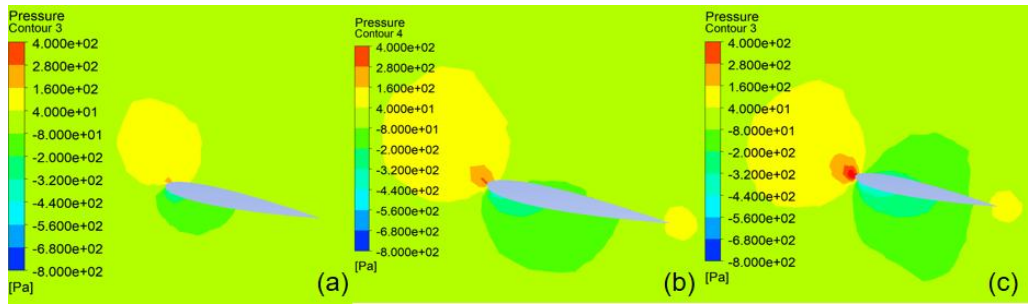

Figure 18. Blade 0.75R section pressure at 12° pitch angle (a) TSR = 2 (b) TSR = 3 (c) TSR = 4

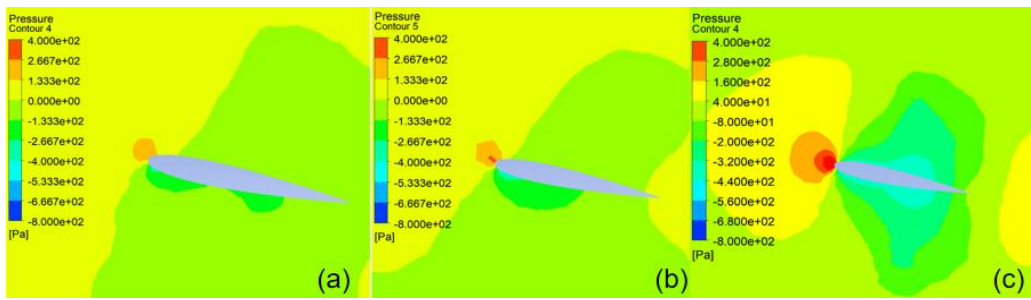

Figure 19. Blade 0.92R section pressure at 12° pitch angle (a) TSR = 2 (b) TSR = 3 (c) TSR = 4

Figures 20-22 present the static pressure cloud at different cross sections for different pitch angles of the blade at the same tip speed ratio. At the position of 0.42R in the blade spreading direction, the static pressure on the leading edge of the blade is small, and with the increase of the pitch angle, the pressure on the leading edge increases obviously, the pressure on the headward surface also increases, and the pressure on the backward surface decreases gradually, and with the blade spreading direction gradually leaning towards the root of the leaf, the pressure on the leading edge of the blade increases gradually, and the pressure on the trailing edge shows a decreasing trend,

and with the increase of the pitch angle, the closer the root of the blade, the phenomenon is about as obvious as it is, therefore. When the tip speed ratio is certain, with the increase of pitch angle, the thrust coefficient of the hydraulic turbine will gradually decrease.

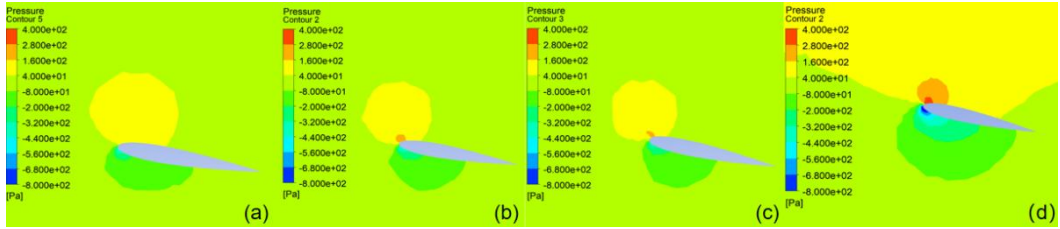

Figure 20. Blade 0.42*R* section pressure at TSR = 3 (a) *β*= 8° (b) *β* = 10° (c) *β* = 12° (d) *β* = 14°

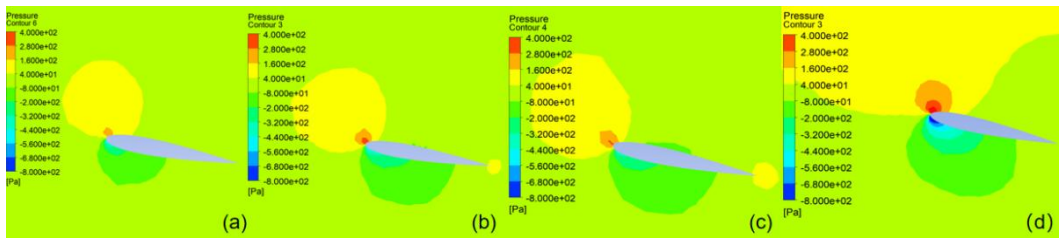

Figure 21. Blade 0.75*R* section pressure at TSR = 3 (a) *β*= 8° (b) *β* = 10° (c) *β* = 12° (d) *β* = 14°

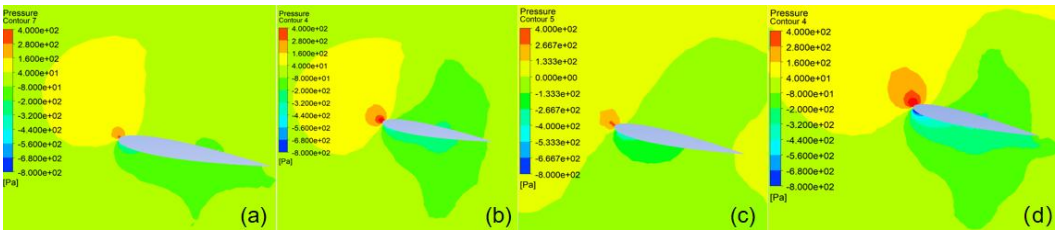

Figure 22. Blade 0.92*R* section pressure at TSR = 3 (a) *β*= 8° (b) *β* = 10° (c) *β* = 12° (d) *β* = 14°

## 5. Experimental Tests

According to the above simulation analysis, the three-bladed hydraulic turbine has the optimal starting performance, which is especially suitable for the utilisation of low speed tidal current resources in most areas of China. Therefore, the number of blades is selected to be 3 and the shaft diameter ratio is 0.7 for experimental study. Figure 23 displays the model experimental test bed. The hydrodynamic performance of the turbine was analysed by measuring the torque, rotational speed and thrust of the turbine at a flow rate of 0.3 m/s.

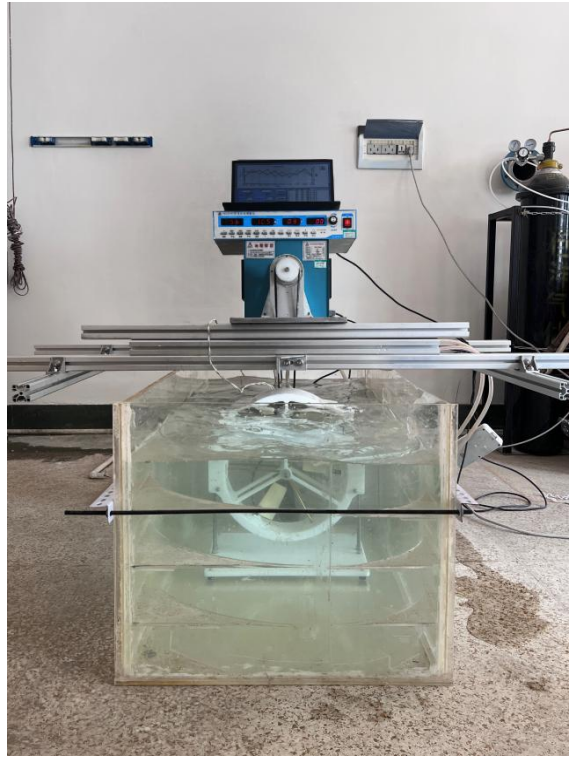

Figure 23. Model experimental testbed

Figure 24 Simulation data of the power coefficient of the turbine at different TSR. From the simulation data and experimental data, it can be seen that the power coefficient shows a trend of increasing and then decreasing with the increase of TSR. When TSR is about 3, the power coefficient is close to the maximum value of 0.303. When TSR is less than 3, the experimental data is close to the simulation value. For TSR values of 3 ~ 6, the experimental values are smaller than the simulated values. This is due to the fact that the simulation is performed under ideal conditions with little energy loss. However, the numerical simulation results reflect the energy harvesting characteristics of the turbine better.

The results of the experimental data compared with the simulated data for the thrust coefficient of the turbine at different TSRs are given in Figure 25. The results show that the variation trend of thrust coefficient is approximately the same in both experimental and numerical simulation environments. When the TSR is 3, the thrust coefficient is 0.75278, which is close to the design condition. When TSR is greater than 3, the simulated value is greater than the experimental value. When TSR is less than 3, the experimental value is similar to the simulated value. The numerical simulation results are basically consistent with the trend of the experimental results and can be used for the prediction of the turbine thrust performance.

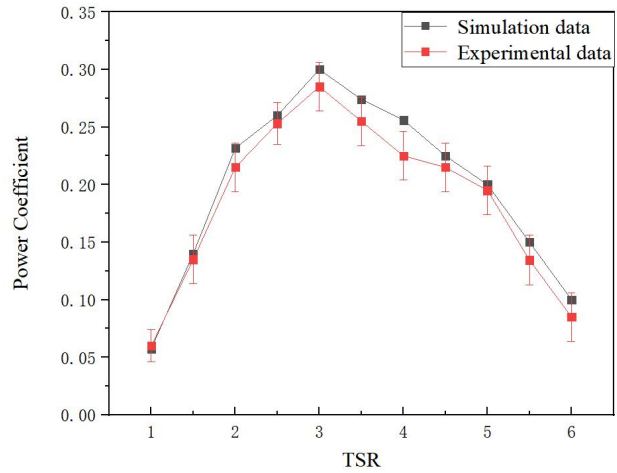

Figure 24. Comparison between experimental and simulated data for power coefficient

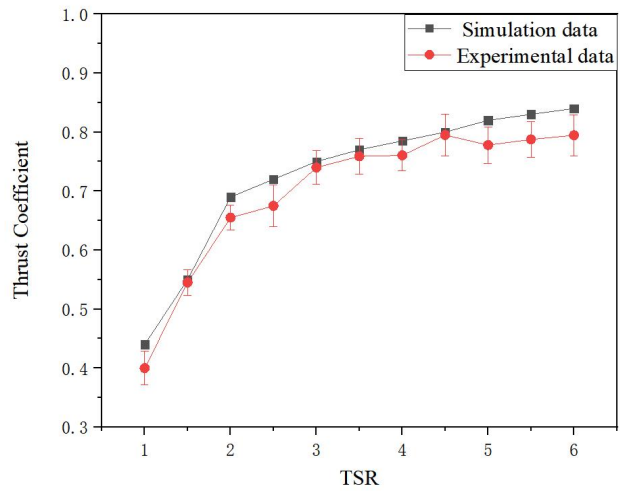

Figure 25. Comparison of experimental and simulation data of thrust coefficient

## 6. Conclusions

The adaptive variable pitch method is applied to the hollow turbine in this paper to achieve easy start-up at low flow rates and more efficient utilisation of bi-directional tidal energy at the same time. The velocity distribution, pressure distribution, energy gain and load characteristics of the hollow turbine are analysed and discussed through simulation calculations, and the simulation results are verified by using flume experiments. The research results demonstrated that:

The effects of different blade numbers and shaft diameter ratios on the energy acquisition and load characteristics of the hydraulic turbine are comparatively analysed by CFD method, and the hollow structure parameters with blade number of 3 and shaft diameter ratio of 0.7 are finally selected, and by analysing the hydrodynamic performances of the hydraulic turbine blades under different pitch angles, taking into account the fact that the energy acquisition coefficient of the impeller is as large as possible and the thrust coefficient is as small as possible, and the optimal pitch angle of 10° is finally selected with the energy acquisition coefficient of 0.303, and the thrust coefficient is about 0.7.

Hollow turbine at the same axial position, the wake velocity increases with the increase of the pitch angle, and when the pitch angle is the same, its wake velocity shows a decreasing and then increasing trend with the increase of the tip-speed ratio, compared with the axial adaptive variable pitch turbine wake velocity will not appear lower flow band, and the recovery of the wake velocity at the same position is better, which will provide reasonable flow velocity requirements for the future arrayed arrangement.

The high pressure of the blade on the flow surface is mainly concentrated in the blade root area, with the increase of the tip speed ratio, the size and range of the pressure on the blade root increase, resulting in a larger torque and thrust acting on the blade, thus increasing the energy acquisition and load characteristics of the turbine; the larger static pressure of the blade is mainly distributed in the leading edge of the blade, and with the blade spreading gradually close to the root of the leaf, the pressure of the leading edge is gradually increased.

The energy acquisition and load characteristics of the hydraulic turbine are tested through the flume experiment, and the obtained experimental results are basically consistent with the simulation results. The power coefficient shows an increasing and then decreasing trend with the increase of the tip speed ratio, and the maximum power coefficient of 0.282 is obtained when the tip speed ratio is 3. The thrust coefficient increases with the increase of the tip speed ratio, and the increase gradually decreases.